# DNA Looping Kinetics Analyzed Using Diffusive Hidden Markov Model

John F. Beausang*, Chiara Zurla†, Carlo Manzo†, David Dunlap††, Laura Finzi†, and Philip C. Nelson*

*Department of Physics and Astronomy, University of Pennsylvania, Philadelphia, Pennsylvania, 19104 USA; Departments of †Physics and ††Cell Biology, Emory University, Atlanta, Georgia, 30322 USA

ABSTRACT   Tethered particle experiments use light microscopy to measure the position of a micrometer-sized bead tethered to a microscope slide via a ~micrometer length polymer, in order to infer the behavior of the invisible polymer. Currently, this method is used to measure rate constants of DNA loop formation and breakdown mediated by repressor protein that binds to the DNA. We report a new technique for measuring these rates using a modified hidden Markov analysis that directly incorporates the diffusive motion of the bead, which is an inherent complication of tethered particle motion because it occurs on a time scale between the sampling frequency and the looping time. We compare looping lifetimes found with our method, which are consistent over a range of sampling frequencies, to those obtained via the traditional threshold-crossing analysis, which vary depending on how the raw data are filtered in the time domain. Our method does not involve such filtering, and so can detect short-lived looping events and sudden changes in looping behavior.

One mechanism for regulating DNA transcription is for a protein to bind to specific operator sites in the DNA sequence, thereby enhancing or diminishing the expression of adjacent genes. In an elaboration of this idea, multiple operators recruit copies of the repressor protein, bind to each other, and bend the DNA into a loop, for example in the lambda system (1). One goal of *in vitro* DNA looping experiments is to determine the rate constants for DNA loop breakdown/formation and gain insight into how the physical process of looping influences the biochemistry of transcription. The purpose of this paper is to present a new diffusive hidden Markov method (DHMM) for determining the looping kinetics from data measured in tethered particle experiments. The main advantage of our method is that by directly incorporating the dynamics of particle diffusion we do not need to filter the raw data. Consequently, DHMM has better time resolution and more consistent results than the traditional threshold-crossing analyses.

The tethered particle method (TPM, Fig. 1) consists of measuring the Brownian motion of a small bead attached to a microscope slide via a short polymer tether, to learn about the tether's behavior (2-6). Our setup uses DIC imaging of a 480 nm diameter polystyrene bead tethered to the slide via a 3477 bp DNA construct containing two sets of three wild-type lambda operator sites separated by 2317 bp, as described previously (7). The ($x,y$) coordinates of up to 6 well-spaced beads are recorded simultaneously with 20 ms time resolution using custom particle tracking software and a CCD camera with a 1 ms shutter to reduce blurring. Bead positions are first recorded for ~10 minutes to ensure uniform behavior. Upon addition of 200 nM cI repressor protein, dynamic exchanges between unlooped and looped tether lengths—consistent with the known construct length and operator spacing—are observed. After recording for 30-60 minutes, the data are corrected for microscope drift and screened for anomalous sticking events using methods described previously (8).

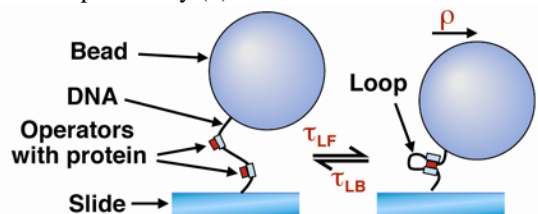

FIGURE 1 Schematic showing DNA loop formation and breakdown in a tethered bead. $\rho$ denotes the plane-projected distance from the attachment point to the bead center.

After drift correction, transitions are clearly visible when the data are plotted as the radial distance from the anchor point $\rho_t = \sqrt{x_t^2 + y_t^2}$, where $t$ is an index indicating which video frame (Fig 2). The equilibrium distributions of $\rho$ are well understood (8, 9) but the large overlap between unlooped and looped distributions at small values of $\rho$ prevents us from unambiguously determining the state of the DNA at particular times—loop formation is not directly observable. Typically, this ambiguity is reduced by filtering $\rho$ (we find the variance over windows of time width $W$); however, the time resolution is then degraded by at least the same amount (6, 7). Filtering helps remove false events at very short times introduced by natural Brownian motion of the bead, but actual events are missed due to the reduced time resolution. Unfortunately, we show below that in our system, looping lifetimes determined by this technique depend strongly on the chosen value of $W$.

Hidden Markov methods (10,11) however, do not require such smoothing. These methods allow for analysis of the unfiltered data, once we overcome one obstacle: In traditional HMM applications, the uninteresting part of the observed signal (or "noise") has no correlations apart from those introduced by the underlying hidden process. Unfortunately, the tethered Brownian motion of our bead has an intrinsic time scale slow compared with our 20 ms sampling frequency, but faster than the looping lifetime. The basic physics is easily reproduced by a particle diffusing in a harmonic potential well (a problem similar to TPM motion), with diffusion constant D = 480,000 nm$^2$/sec and spring constant κ = 0.06 x 10$^{-3}$ pN/nm obtained from fits to the autocorrelation of the measured position, the characteristic decay time $\tau_D = k_BT/D\kappa$ is ~140 ms (12). This diffusive motion not only prevents efficient filtering, but also the direct application of traditional hidden Markov methods to TPM.

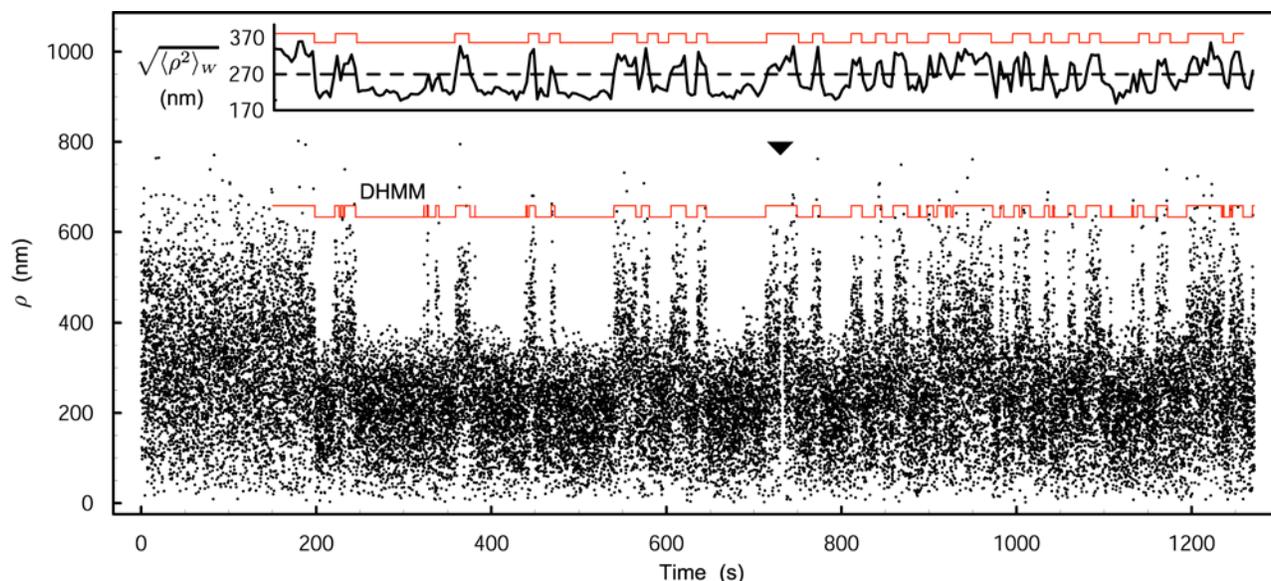

FIGURE 2 Dynamic looping seen in tethered particle time series data. The region from time 200—1200 s shows dynamic looping behavior and was used in the subsequent analysis. The ▼ indicates a brief 4.5 s sticking event that was omitted from the analysis by concatenating the drift-corrected data. The inset shows the corresponding filtered time series with window size $W$ = 4 s. Red lines are inferred transition sequences for DHMM and threshold-crossing analyses, respectively. Dashed line = 270 nm threshold.

More precisely, standard HMM supposes that an observed signal reflects two processes (10): A hidden process that generates a time series $\{q_t\}$ according to an autonomous Markov process with some time-step distribution $D(q_{t+1}|q_t)$, and an observed signal $\{r_t\}$ that at each instant $t$, is drawn from a probability distribution $P(r_t|q_t)$, which depends only on the current value of $q_t$. This framework is appropriate for the case where $q_t$ is the internal state of an ion channel and $r_t$ is the instantaneous current through the channel. We might be tempted to apply it to our case as well, letting $q_t$ denote the looping state of our DNA tether and $r_t = (x_t, y_t)$. But the ability to form a loop depends on the location of the bead: For example, if the bead is too far from the attachment point, then loop formation is impossible until the bead has wandered closer, invalidating the assumption made in standard HMM.

Moreover, the next bead location $r_{t+1}$ depends not only on the present looping state, but also on the present bead location (if the chosen time step is not much longer than the bead diffusion time). For both of these reasons, we must modify the usual formulation of HMM.

To find the required modification, we first note when no cI protein is present, the bead executes tethered Brownian motion, and this motion is itself a Markov process: The bead's displacement $r_{t+1}$ depends only on $r_t$, not on earlier positions. We extracted the "unlooped" probability distribution for the next position, $D_{un}(r_{t+1}|r_t)$, from observed time series in a control experiment. Then we found the analogous distribution $D_{loop}(r_{t+1}|r_t)$ for permanently looped tethers. Our two distributions $D_{un}$ and $D_{loop}$ were thus determined phenomenologically, with no attempt to model the dynamical details of tethered Brownian motion near a wall. As functions of $r_{t+1}$, the distributions $D_{un}$ and $D_{loop}$ are both roughly 2D gaussians centered about a point that depends on $r_t$, with widths that reflect the random excursions of Brownian motion in one time step. We checked that simulating Markov processes with these distributions gave good agreement with the control data, both for the probability distributions of the radial distance $\rho$, and for the autocorrelation functions of $x$ and $y$, two nontrivial consistency checks on our data and theory.

In order to incorporate the hidden state dependence, we constructed a heuristic joint distribution function $D_{DHMM}(q_{t+1},r_{t+1}|q_t,r_t)$, the probability of observing $q_{t+1},r_{t+1}$ given $q_t,r_t$, as follows. If the DNA is initially unlooped ($q_t$=1) and $\rho_t$ is too large to permit loop formation, then the DNA must remain unlooped in the final state: $D_{DHMM}=D_{un}(r_{t+1}|r_t)$ for $q_{t+1}$=1 and $D_{DHMM}$=0 for $q_{t+1}$=2.

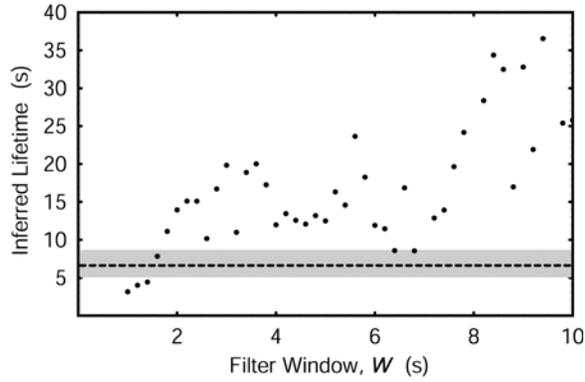

**FIGURE 3** Lifetime results for the threshold method for DNA loop formation (points) of the measured data shown in Fig. 2, as a function of filter window W. Dwell times between threshold crossings in the unlooped/ looped state longer than twice the filter dead time, i.e. 2W, were fit to a histogram to determine $\tau_{LF}$ and $\tau_{LB}$ (13). Our DHMM method uses no window; its result is shown as the dashed line. The shaded region corresponds to the error estimate discussed in the text.

However, if $q_t = 1$ and $\rho_t$ is less than the maximum excursion observed for beads with a permanently looped tether (observed in a separate experiment and verified via Monte Carlo simulation (9)), then both final states are allowed, and we take $D_{DHMM}=(1-\Delta t/\tau_{LF}) \cdot D_{un}(r_{t+1}|r_t)$ for $q_{t+1}=1$ and $D_{DHMM}=\Delta t/\tau_{LF} \cdot D_{loop}(r_{t+1}|r_t)$ for $q_{t+1}=2$. The rate constant $1/\tau_{LF}$ is a parameter of the model, the probability per time to form a loop when permitted. A similar construction gives the case when the DNA is inially looped ($q_t=2$), in terms of a second unknown rate constant $1/\tau_{LB}$ for loop breakdown.

We repeated the above calculation for all pairs of data points and summed over all possible sets of the hidden variables $q_t$ (10), resulting in a likelihood function:

$$P(\mathbf{r}_N, ... \mathbf{r}_1) = \sum_{\{q_i\}} D_{DHMM}(\mathbf{r}_N, q_N | \mathbf{r}_{N-1}, q_{N-1}) \times ...$$
$$\times D_{DHMM}(\mathbf{r}_2, q_2 | \mathbf{r}_1, q_1) P(\mathbf{r}_1, q_1)$$

with two unknown fit parameters, $\tau_{LF}$ and $\tau_{LB}$, which are the quantities of interest to us. We evaluated the likelihood for various values of the parameters, expressing it as a 2D surface on a logarithmically spaced grid. The resulting surface is smooth, so the peak likelihood can be determined by fitting a 2D quadratic in the neighborhood of the optimum lifetimes, including error bars corresponding to a range of lifetimes enclosing the maximum with 97% confidence. A coumpter code implementing our algorithm is available as supplementary material.

We tested our DHMM by analyzing multiple data subsets obtained by thinning the data, by either a factor of two ($\Delta t = 40$ ms) or four ($\Delta t = 80$ ms). All computed lifetimes and cross validation of the likelihoods between independent data subsets agreed within uncertainty. To test the algorithm further, we generated a Monte Carlo simulation of the 40-ms looping data, assuming the values of $\tau_{LB}$ and $\tau_{LF}$ determined from the experimental data. Then we applied our DHMM method to the simulated data, and checked that it again found the known values and that the event detection corresponded to the time series of the hidden looping transitions (which were known in the simulated data). In contrast to these consistent results, we found that the threshold-crossing method resulted in a lifetime that depends on the filter window size W; see Fig 3 where for simplicity only $\tau_{LF}$ is shown. (Additional tests and further mathematical details of the method will be discussed elsewhere.)

We have developed a new method for assessing DNA looping rates from data obtained by the tethered particle method. We tested it on actual and simulated data and determined lifetimes that were independent of sampling frequency. DHMM should improve TPM as a quantitative tool, providing more consistent results with improved time resolution compared to the threshold-crossing method.

### ACKNOWLEDGEMENTS

We thank Rob Phillips for recommending HMM methods for this problem and Seth Blumberg, Lin Han, Randall Kamien, and Liam Paninski for useful discussions. We thank Yale Goldman and the anonymous referee for critical suggestions on the manuscript. This work was supported in part by the Human Frontier Science Programme and NSF grants DGE-0221664, DMR04-25780, and DMR-0404674. LF and PN acknowledge the hospitality of the Kavli Institute for Theoretical Physics, supported in part by the National Science Foundation under Grant PHY99-07949.

### REFERENCES

1. Ptashne, M. 2004. A Genetic Switch. CSHL Press, New York.
2. Schafer, D. A., J. Gelles, M. P. Sheetz, and R. Landick. 1991. Transcription by single molecules of RNA polymerase observed by light microscopy. Nature 352:444-448.
3. Finzi, L., and J. Gelles. 1995. Measurement of Lactose Repressor-Mediated Loop Formation and Breakdown in Single DNA-Molecules. Science 267:378-380.
4. Vanzi, F., S. Vladimirov, C. R. Knudsen, Y. E. Goldman, and B. S. Cooperman. 2003. Protein synthesis by single ribosomes. RNA 9:1174-1179.
5. van den Broek, B., F. Vanzi, D. Normanno, F. S. Pavone, and G. J. L. Wuite. 2006. Real-time observation of DNA looping dynamics of type IIE restriction enzymes NaeI and NarI. Nucl Acids Res 34:167-174.
6. Vanzi, F., C. Broggio, L. Sacconi, and F. S. Pavone. 2006. Lac repressor hinge flexibility and DNA looping. Nucleic Acids Res 34:3409-3420.
7. Zurla, C., A. Franzini, G. Galli, D. D. Dunlap, D. E. A. Lewis, S. Adhya, and L. Finzi. 2006. Novel tethered particle motion analysis of CI protein-mediated DNA looping in the regulation of bacteriophage lambda. J Phys: Condens Matter 18:S225-S234.
8. Nelson, P. C., C. Zurla, D. Brogioli, J. F. Beausang, L. Finzi, and D. Dunlap. 2006. Tethered particle motion as a diagnostic of DNA tether length. J. Phys. Chem. B 110:17260-17267.
9. Segall, D. E., P. C. Nelson, and R. Phillips. 2006. Volume-exclusion effects in tethered-particle experiments. Phys Rev Lett 96:088306.
10. Rabiner, L. R. 1989. A tutorial on hidden Markov models and selected applications in speech recognition. Proc. IEEE 77:257-286.
11. McKinney, S. A., C. Joo, and T. Ha. 2006. Analysis of single-molecule FRET trajectories using hidden Markov modeling. Biophys J 91:1941-1951.
12. Howard, J. 2001. Mechanics of Motor Proteins and the Cytoskeleton. Sinauer Assoc., Sunderland, MA.
13. Colquhoun, D., and F. J. Sigworth. 1983. Fitting and statistical analysis of single-channel records. In Single-Channel Recording. B. Sakmann, and E. Neher, editors. Plenum, New York. 191-263.